\documentstyle[amssymb,epic,eepic,epsffrag,psfrag,graphicx,preprint,aps]{revtex}

\newcommand{\bm}[1]{\mbox{\boldmath$#1$}}

\newcommand{\pard}[2]{\frac{d{#1}}{d{#2}}}

\begin{document}
\title{Motion of Quantized Vortices as Elementary Objects}
\author{Uwe R. Fischer\cite{email}}
\address{
Lehrstuhl f\"ur Theoretische Festk\"orperphysik,
Institut f\"ur Theoretische Physik,
Universit\"at T\"ubingen, 
Auf der Morgenstelle 14, D-72076 T\"ubingen, Germany}

\date{\today}
\maketitle


\begin{abstract}                
The general local, nondissipative 
equations of motion for a quantized vortex 
moving in an uncharged laboratory superfluid are derived 
from a relativistic, co-ordinate invariant framework, having vortices 
as its elementary objects in the form of stable topological 
excitations. This derivation is carried out 
for a pure superfluid with isotropic gap at the 
absolute zero of temperature, on the level of a hydrodynamic, 
collective co-ordinate description.   
In the formalism, we use as fundamental 
ingredients that particle number as well as vorticity are conserved, and 
that the fluid is perfect.
No assumptions are involved as regards the dynamical behaviour of the order
parameter. 
The interaction of the vortex with the background fluid, 
representing the Magnus force, and with itself {\it via} phonons, 
giving rise to the hydrodynamic vortex mass, are separated. For a 
description of the motion of the vortex in a dense laboratory 
superfluid like helium II,  two limits have to be considered: 
The nonrelativistic limit for the superfluid background is taken, and
the motion of the vortex is restricted to velocities much less 
than the speed of sound. The canonical structure of vortex motion 
in terms of the collective co-ordinate is used for the quantization of this
motion.  
\end{abstract}

\


\


\newpage
\section{Introduction}
Vortices are the fundamental line-like excitations of a fluid. Their motion
governs a large number of properties of the fluid under given external 
conditions. Superfluids 
are peculiar in that the vortices can exist only in certain classes, 
distinguished by integers. This fact that vorticity can only arise 
quantized, results in far-reaching consequences for the character 
of vortex generation and motion in quantum fluids, as compared to simple
perfect fluids.   
The aim of the present investigation lies in the 
derivation of the general local 
equations of motion for a quantized vortex in a 
laboratory superfluid at absolute zero,
thereby starting from a minimum of very basic assumptions about the 
co-ordinate invariant nature of the hydrodynamic conservation laws in a 
relativistic perfect fluid and the Magnus force acting on the vortex.  
In particular, no assumptions will be made with respect to the underlying 
dynamics of the 
order parameter of the superfluid.   

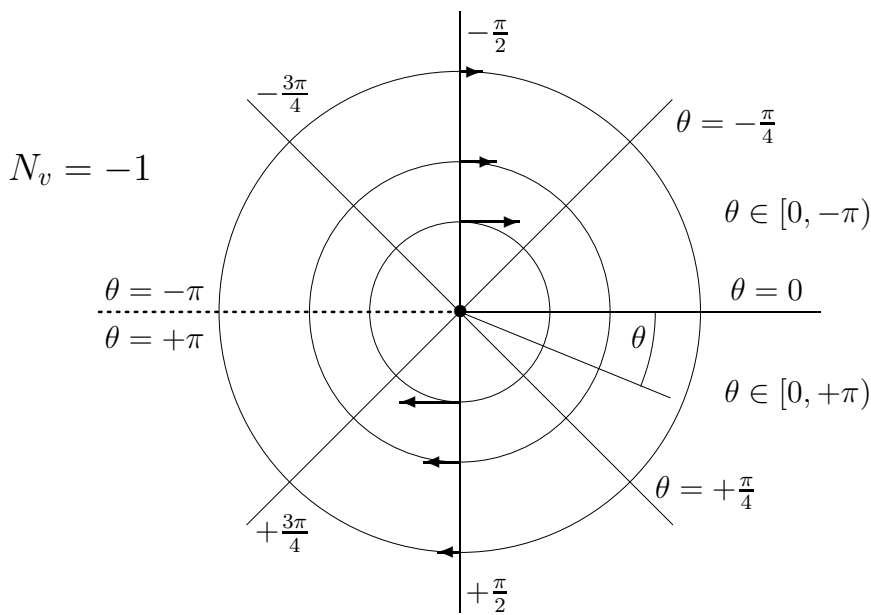
\begin{figure}[hbt]
\begin{center}

  \unitlength=0.8mm
  \begin{picture}(100,100)(-50,-50)
\put(-1.2,-1.2){$\bullet$}
 \put(0,0){\circle{30}}
\put(0,0){\circle{50}}  
\put(0,0){\circle{80}}
\put(0,0){\line(1,0){60}}
\thicklines
\put(0,0){\dottedline{1.5}(-60,0)(0,0)}
\thinlines
\put(-35.4,-35.4){\line(1,1){70.71}}
\put(35.4,-35.4){\line(-1,1){70.71}}
\put(0,-50){\line(0,1){100}}
\put(0,0){\arc{65}{0}{0.39}}
\put(0,0){\line(100,-41){35}}
\put(28.5,-6){\makebox(0,0)[lb]{$\displaystyle \theta$}}
\put(45,2){\makebox(0,0)[lb]{$\displaystyle \theta=0$}}
\put(-75,22){\makebox(0,0)[lb]{\large $N_v=-1$}}
\put(36,30){\makebox(0,0)[lb]{$\textstyle \theta=-\frac\pi 4$}}
\put(1,45){\makebox(0,0)[lb]{$\textstyle  -\frac\pi 2 $}}
\put(-34,35){\makebox(0,0)[lb]{$\textstyle -\frac{3\pi}4$}}
\put(-59,2){\makebox(0,0)[lb]{$\displaystyle \theta=-\pi $}}
\put(-59,-6){\makebox(0,0)[lb]{$\displaystyle \theta= +\pi $}}
\put(-34,-38){\makebox(0,0)[lb]{$\textstyle +\frac{3\pi}4$}}
\put(1,-48){\makebox(0,0)[lb]{$\textstyle +\frac\pi 2$}}
\put(32.5,-31){\makebox(0,0)[lb]{$\textstyle \theta=+\frac\pi 4$}}
\put(44,15){\makebox(0,0)[lb]{$\displaystyle \theta\in [0,-\pi)$}}
\put(44,-15){\makebox(0,0)[lb]{$\displaystyle \theta\in [0,+\pi)$}}
\thicklines
\put(0,40){\vector(1,0){3.75}}
\put(0,25){\vector(1,0){6}}
\put(0,15){\vector(1,0){10}}
\put(0,-40){\vector(-1,0){3.75}}
\put(0,-25){\vector(-1,0){6}}
\put(0,-15){\vector(-1,0){10}}
\end{picture}
\end{center} 
\caption{\label{vortexisolated} A rectilinear 
quantized vortex with {\em negative} winding number $N_v=-1$. 
The superflow in the $x$-$y$ plane, indicated by the arrows, is in
{\em clockwise} direction along the gradient of $\theta$. For a vortex
with positive winding number the superflow is anticlockwise,
correspondingly. The branch cut is chosen to be at $\theta=\pm\pi$ and
the circulation vector is pointing into the figure.}
\end{figure}

The most fundamental property of a vortex living in a 
superfluid is its topological stability. 
It is a topological line defect
belonging to the first homotopy group $\Pi_1(G/H)$, which 
contains the equivalence classes of loops around a singularity: 
The circle $S^1$ in real space is mapped onto the coset space $G/H$, and 
the equivalence classes identified. Here,  
$G$ is the symmetry group of the system under consideration, and $H$ is 
the little group, a subgroup of $G$, characterizing the partial symmetry 
of the system which remains after the symmetry of the larger group $G$ 
has been broken.
This coset space represents the manifold of the degenerate vacuum
states after spontaneous symmetry breakdown, and can be  identified with 
the order parameter. 
In this paper, we will deal exclusively with a very 
simple such coset space, namely the circle $S^1$ 
(the group $G$ is U(1) and $H$ is the identity), 
so that the mapping in question is between the circle in 
ordinary and that in order parameter space. The winding number
$N_v\in {\mathbb{Z}}=\Pi_1(S^1)$
classifying the paths around the line equivalent to each other, 
indicates the number of times one is `winding' around the singular line in
order parameter space, if one is going once around it in real space 
(see Fig. \ref{vortexisolated}).   
The complex order parameter of a spontaneously broken U(1) symmetry, 
with phase $\theta$,
then gives rise to the quantization of the 
nonrelativistic superfluid circulation $\Gamma_s$
in units of $\kappa = h/ m $, 
\begin{equation}
\Gamma_s=\oint
{\vec v}_s\cdot d{\vec s}
=\frac\hbar {m} \oint
\nabla\theta\cdot d{\vec s}
=N_v \cdot \frac h {m} \equiv N_v\cdot \kappa
\, ,\label{Gammas}
\end{equation}
provided the superfluid velocity is identified as
${\vec v}_s = (\hbar/m)\nabla \theta$. 
For a vortex in the superfluid helium II, the superfluid particle's mass  
$m=6.64\cdot 10^{-27}\,$kg, so that $\kappa\simeq
10^{-7}\,$m$^2$/s. 
The definition in (\ref{Gammas}) is the classical {\em kinematical} 
definition of circulation \cite{lamb,milne-t}.  
We will see below that a relativistically 
invariant, {\em dynamical} 
definition has to use the line integral of the momentum, 
rather than that of the velocity, to gain truly invariant meaning. 
However, in any case, be it relativistic or nonrelativistic, 
the stability (and very definability) 
of the quantized vortex as a fundamental object able to persist 
while moving in the superfluid stems essentially from its 
topological properties. The superfluid 
cannot be continously deformed into a vortex-free state.
On the other hand, this feature of topological stability has its
limits if large quantum statistical fluctuations of the 
order parameter become important. In the dense, unpaired  superfluid 
helium II, they become large on the same coherence length 
scales, on which the order parameter modulus varies, of the order
of the inter-particle distance.  
Hence a vortex, having curvature scales of the
singular line, which are of order the inter-particle distance, 
is not definable topologically 
and thus does not exist as a well-defined line defect 
in order parameter space. On these scales, it has to be defined using 
the full quantum many-body structure of the dense superfluid and the 
corresponding vortex wave functions.  

This hydrodynamic level of accuracy can only be afforded by the fully 
relativistic description of the dense liquid we present in the section which 
follows, too, because the same problem of incorporating 
microscopic many-body structure excitations into the description remains, 
which exists already in the nonrelativistic superfluid. 
A hydrodynamic formalism
is thus a necessity to describe vortex motion properly, as long as this 
problem of microscopic dynamics has not been solved. 
The covariant treatment, in turn, provides structural 
insights which are not given by more conventional treatments.      
\section{Duality of Vortices and Charged Strings}\label{secdual}
We establish in this section the properties of a vortex as a
stringlike fundamental object. For this purpose, 
the means of dual transformation will be instrumental 
(where {\it dual} is meant 
in the sense of being `equivalent as a physical system'). We will
introduce the duality of interest here 
in its relativistic context (established in
Refs. \cite{kalb,lundregge}, elaborations in various directions
can be found in Refs. 
\cite{davis1}--\cite{langlois2}),  
subsequently reducing it to its 
nonrelativistic limit.

Conventions are that the velocity of light $c$ is set equal to unity and 
the signature $\eta_{\mu\nu} = {\rm diag}\,(-1,1,1,1)$ 
of the metric is being employed.
Greek indices mean spacetime indices and take the values
$0,1,2,3$. 
The sign convention we use for the Levi-Civita pseudotensor 
is $\epsilon^{0123}=+1=-\epsilon_{0123}$ in a 
Lorentz frame.
The dual of a $p$-form ${\bm f}$ 
in four-dimensional spacetime has the contravariant components 
$^*\!f^{\beta_1\ldots \beta_{4-p}}= (1/p!)f_{\alpha_1\ldots\alpha_p}
\epsilon^{\alpha_1\ldots \alpha_p \beta_1 \ldots \beta_{4-p}}$.


The hydrodynamic conservation law 
we will use as a first basic ingredient of our theory 
for an uncharged relativistic fluid is that of particle number. 
The most familiar 
mathematical form of nonrelativistic number conservation is provided
by $\partial_t \rho + {\rm div}({\rho \vec v}_s)=0$.  
This is expressed covariantly, using $p-$forms \cite{MTW},  as   
\begin{equation}
 {\bm d}\wedge\!^*\!{\bm j}=0 
\qquad ({\bm d}\equiv {\bm d}x^\alpha\partial_\alpha)\, .
\label{J}
\end{equation}
The number current one-form is defined 
${\bm j}\equiv \rho {\bm u}=\rho u_\alpha {\bm d} x^\alpha$, 
where ${\bm d} x^\mu$ is a one-form basis dual to a co-ordinate basis 
and $\rho$ the rest frame number density.
We omit the subscript $s$ from the relativistic quantity 
$\bm u$, whose normalisation will be required by its interpretation as a 
four-velocity to be $u^\alpha u_\alpha= -1$. 

The momentum of a perfect fluid particle 
is defined as the one-form $\bm p$, with 
components \cite{carterkhalat}
\begin{equation}\label{pmudef}
p_\alpha =\mu u_\alpha\,.
\end{equation} 
The chemical potential $\mu = \partial \epsilon/\partial \rho $, where
$\epsilon$ is the rest frame energy density, then plays the role of an
effective rest mass in a Hamiltonian quadratic in the four-momentum.  
The momentum and number current density one-forms are thus related through
\begin{equation}
\frac\rho\mu\, p_\alpha = \frac K{\hbar^2}\, p_\alpha =  j_\alpha\,.
\end{equation} 
The quantity $K=\hbar^2 (\rho/\mu)$ is the stiffness 
coefficient against variations of the order parameter phase
(a quantity to be defined below) in the free energy, 
cf. \cite{carterkhalat} and section 3.1 in \cite{grishabook}. 
The speed of sound $c_s$ is related to the quantities introduced in the 
above, for a barotropic fluid \cite{langlois1}, as $c_s^2 =
d(\ln \mu) /d(\ln \rho)
= (K/\hbar^2) d^2 \epsilon /d \rho^2$.
    
We define the vorticity as the dual of the exterior derivative 
of $\bm p$:
\begin{equation}\label{omegadef}
{\bm \omega} = {^*\!{\bm d}}\wedge{\bm p} \quad \Leftrightarrow \quad 
^*\!{\bm \omega} = - {\bm d}\wedge{\bm p}\,.
\end{equation}
The 
vorticity thus defined is 
required to be conserved, 
in the form
\begin{equation}\label{omegaconserv}
{\bm d}\wedge^*\! {\bm \omega}=0\quad \Leftrightarrow \quad
{\bm d}\wedge {\bm d}\wedge{\bm p}=0\,.
\end{equation}

Outside the cores of the vortices, where the quantity  
vorticity $\bm \omega$ defined above equals zero,  
the momentum of the fluid always remains proportional to 
the (exterior) derivative of a scalar $\theta$, which we identify with 
the phase of the U(1) order parameter pertaining to the superfluid:
\begin{equation}\label{palphatheta}
p_\alpha = \hbar \partial_\alpha \theta\,.  
\end{equation}  
The circulation is in the relativistic case defined as the integral 
of the momentum. This choice of definition stems from the fact that 
a particle's rest mass, occuring in the  nonrelativistic definition of 
the circulation (\ref{Gammas}), is an undefined quantity in a 
fully relativistic, dense superfluid. The proper 
circulation thus reads 
\begin{equation}\label{gammas}
\gamma_s= \oint p_\mu dx^\mu = N_v h \,,
\end{equation}
and is quantized into multiples of Planck's quantum of action: 
The constant of proportionality $\hbar=h/2\pi$ in (\ref{palphatheta}) 
stems from the covariant Bohr-Sommerfeld 
quantization of $\gamma_s$, {\it i.e.} from the fact 
that we require ${\bm p}$ to be a quantum variable. 
It may be worthwhile to point out in this context 
that particularly striking evidence for 
the fundamentality of the {\em one-form quantity}\, momentum,  
${\bm p}= p_\mu {\bm d}x^\mu$,  is given by
consideration of superfluid rotation in the presence of gravity 
\cite{kirzhnitsyudin}, {\it e.g.}, in neutron stars. 
Superfluid irrotationality 
is equivalent, for axial symmetry, to $p_\phi =0$, whereas the contravariant 
axial component $p^\phi$ and thus the contravariant axial  
velocity do not have to vanish.    

The fact that quantization of circulation as a line integral of momentum 
is crucial for a proper understanding of the nature of the line defect 
vortex is not connected to the fact that the superfluid is relativistic or
nonrelativistic. The point is rather that the vortex is in general 
nothing but a singular line of zeroes in the order parameter manifold, 
designated by a single number $N_v$, which represents the order of the pole. 
This fundamental, intrinsic  
property of the vortex is not related to a quantity like mass, characterizing 
the matter with which it interacts, but is independent of matter properties. 

The normalisation of $u_\mu$, by using the metric and writing 
$u^\alpha u_\alpha= -1$, is promoting $u^\mu$ into a four-velocity.
It leads, with (\ref{pmudef}) and (\ref{palphatheta}), to the relation 
\begin{equation}\label{mutheta}
\mu = \hbar\left(-\partial_\mu \theta \,\partial^\mu \theta\right)^{1/2}\,
\end{equation}
between the phase and the chemical potential.  
This is a relativistically covariant version of the 
Josephson equation $\hbar\dot\theta = -\left(\mu + \frac12 m{\vec v}_s^2
\right)$, 
familiar from nonrelativistic condensed matter physics \cite{anderson}.  
The Josephson equation expresses, generally, the conjugateness
of the two canonical variables phase 
and number density in the hydrodynamic limit. 

The dual of the 
current ${\bm j}=j_\mu {\bm d} x^\mu$ 
is a 3-form $^*\!{\bm j}= 
j^\mu \epsilon_{\mu\nu\alpha\beta}\, {\bm d} x^\nu \wedge {\bm d}
x^\alpha \wedge {\bm d} x^\beta$, 
which is, according to the conservation law above,
closed.
We define the {\em field strength} $\bm H$ by 
\begin{equation}\label{Hdef}
^*\!{\bm j}= 
\bm H\,.
\end{equation} 
The field $H_{\mu\nu\alpha}$ is totally antisymmetric in its
three indices and has, by definition, only four independent
components. 
In a simply connected region, $\bm H$ is exact, {\it i.e.}, it is the exterior 
derivative of a {\em gauge} 2-form $\bm b\, = b_{\mu\nu}\, {\bm d} x^\mu
\wedge {\bm d} x^\nu $: 
\begin{eqnarray}
H_{\alpha\beta\gamma} =\partial_\alpha b_{\beta\gamma}
+ \partial_\gamma b_{\alpha\beta}+ \partial_\beta b_{\gamma\alpha}
\,,\label{dualtranscomp}
\end{eqnarray}
The field strength
$\bm H$ is invariant under gauge transformations ${\bm b}
\rightarrow {\bm b}+{\bm d}\wedge{\bm \Lambda}$, where ${\bm \Lambda}
=\Lambda_\alpha {\bm d} x^\alpha $ is an arbitrary 1-form. 
Thus
\begin{equation}
b_{\mu\nu}\rightarrow 
b_{\mu\nu}+\partial_\mu \Lambda_\nu -\partial_\nu\Lambda_\mu 
\quad \Rightarrow \quad 
H_{\mu\nu\alpha}\rightarrow H_{\mu\nu\alpha}  \,.
\end{equation}
We then have the following sequence of relations, which define the 
correspondences of $\rho,\mu,\theta$ and $\bm b$:
\begin{eqnarray}\label{dualtrans}
^*\!{\bm j}= \frac\rho\mu\, {^*\!{\bm p}}
=\hbar\frac\rho\mu\,^*\!{\bm d}\theta
=
{\bm d}\wedge{\bm b}=
{\bm H},\label{bdef} 
\end{eqnarray} 
The dual 
transformation in the conventional sense
\cite{kalb,lundregge} obtains 
if we neglect any possible variations of $\rho/\mu = K/ \hbar^2$ 
in the spacetime domain of interest. 
Then, ${\bm b}$ has only one degree of freedom
corresponding to the order parameter phase $\theta$. 

From the equation 
${\bm H}=(\rho/\mu){}^*\!{\bm p}$, it follows that 
\begin{equation}\label{rho0rhoH}
^*\!{\bm d}\wedge \left(\frac{\mu}{\rho}\,{^*\!{\bm H}}\right)=
\,{^*\!{\bm d}}\wedge \left(\frac{\hbar^2}{K}\,{^*\!{\bm H}}\right)
=\,{^*\!{\bm d}}\wedge{\bm p}
= {\bm \omega}\,,
\end{equation}
which is the general 
field equation for the motion 
of vortices with conserved vorticity $\bm \omega$. 
It is a field equation analogous to the inhomogeneous Maxwell
equation. 
In case that the stiffness ratio of density and chemical potential
is a constant $K_0$ 
in space and time, and in Lorenz gauge 
$ ^*\!{\bm d}\wedge\!{^* {\bm b}}=0$, 
the wave equation of the gauge field takes 
the familiar form 
\begin{equation}
\square\, {\bm b}=-(K_0/\hbar^2)\, {\bm \omega}\,,
\end{equation}
where the 
d'Alembertian $\square \equiv \partial_\mu\partial^\mu$.
The homogeneous Maxwell equation for strings is, by definition, 
for any value of $\rho/\mu$,
\begin{equation}\label{dH0}
{\bm d}\,\wedge {\bm H}=0 \qquad (\mbox{arbitrary $\rho/\mu=K/\hbar^2 $})\,.
\end{equation}  

The advantage of using $\bm b$ over $\theta$
consists in the fact that it can be chosen to be single-valued,
which is to be contrasted with the necessary multivaluedness of
$\theta$ in the presence of a vortex.
The physical significance of $\bm b$ is that it 
represents a generalization of the stream function concept encountered 
in classical nonrelativistic hydrodynamics \cite{milne-t}. Namely, 
integrating 
\begin{equation}
\int_{^{(2)}\partial\Omega}b_{|\mu\nu|}{\bm d}x^\mu \wedge {\bm d}x^\nu
=\int_{^{(3)}\Omega}
\rho u^\lambda\,\epsilon_{\lambda|\mu\nu\alpha|}{\bm d}x^\mu\wedge
{\bm d}x^\nu\wedge{\bm d}x^\alpha\,,\label{bstreamf}
\end{equation}
we see that the integral of $\bm b$ over any 2-surface
$^{(2)}\partial\Omega$ enclosing 
a 3-surface $^{(3)}\Omega$ in spacetime is given by the superflow flux 
through the 3-surface (the symbol $|\cdots |$ indicates ordering
of the indices contained between the vertical lines
with increasing numerical value). The 2-form quantity $\bm b $ can thus be
called a generalized 
{\it stream tensor}. The relation (\ref{bstreamf}) expresses a
generalization of the usual concept of stream function (for which the 
incompressibility condition $\rho=\rho_0$ holds), to flows for which the 
density can vary.

In particular, in the nonrelativistic case, which has 
$u^0= 1$, setting 
$\rho= \rho_0$, 
\begin{equation}\label{bintVol}
\frac 12 \int_{^{(2)}\partial\Omega}
b_{ji}\,{\bm d}x^i \wedge {\bm d}x^j
= \rho_0 V_{^{(3)}\Omega}\,,
\end{equation}
that is, the specified 
surface integral of the purely spatial part of $\bm b$ is the
spatial volume enclosed by the surface $^{(2)}\partial\Omega$ 
times the density $\rho_0$, and thus the bulk number of particles
contained in the volume $^{(3)}\Omega$.

The superfluid we will be dealing with later in section \ref{canonstructure} 
is nonrelativistic.  
We thus have to rewrite (\ref{bdef}) 
in a form having only Galilean invariance.
This is accomplished by writing (\ref{bdef}) 
(with $\rho=\rho_0$) for spatial and temporal
indices separately, 
\begin{eqnarray}
\rho_0 {u^0}\epsilon_{0ijk} & = & \partial_k b_{ij}+\partial_j
b_{ki}+\partial_i b_{jk}=H_{ijk}\nonumber\\
\rho_0 u^i{}\epsilon_{i0jk} & = & \partial_k b_{0j}-\partial_j b_{0k}
+\frac 1c \partial_t b_{jk}=H_{0jk}\,.  
\end{eqnarray}
and taking the Gali\-lean limit $c\rightarrow \infty$ ($u^0\rightarrow
c$, $u^i\rightarrow v_s^i$\,; 
we have temporarily reinstated the velocity of light for this purpose).   
We then get the relations (in which $\epsilon_{0ijk}=-\epsilon_{ijk}=
-\sqrt g\, n_{ijk}$):
\begin{eqnarray}\label{bijk}
-\sqrt g\, n_{ijk}\rho_0 = 
\partial_k b_{ij}+\partial_j b_{ki}+\partial_i b_{jk}\,, \\
\label{stream}
\sqrt g \, n_{ijk}\,v_s^i =  \partial_k\psi_j -\partial_j \psi_k\,,
\end{eqnarray}
where the  vectorial version of the usual stream
function \cite{milne-t} is defined through
\begin{equation}\label{psiidef}
\psi_i \equiv b_{0i}/\rho_0\,.
\end{equation}
The determinant of the spatial co-ordinate system we are using 
is designated $g$, and $n_{ijk}=n_{[ijk]}=\pm 1$ is the unit 
antisymmetric symbol.

\subsection{The Magnus force}\label{subsectionMagnus}
The fundamental force acting on a superfluid vortex at the absolute
zero of temperature is the Magnus force. 
This force is now shown to be equivalent to a stringy generalization of 
the Lorentz force. Given this identification, we will be 
able to write down electrodynamic correspondences in the
next subsection. 

The Magnus force acting on a vortex in a nonrelativistic superfluid
has the standard form
\begin{eqnarray}
{\vec F}_M & = & 
m\rho_0  {\vec \Gamma}_s \times \left( \dot {\vec X}-{\vec v}_s\right)
\nonumber\\
& = & \gamma_s\rho_0 {\vec X}'\times \left( \dot {\vec X}-{\vec v}_s\right)
\label{magnusforce} \,,
\end{eqnarray}
where ${\vec v}_s$ is the superflow velocity far from 
the vortex center located at $X^i(t,\sigma)$ ({\it i.e.} the flow field at
the vortex location without the contribution of the vortex itself),  
and  $\sigma$ is 
the arc length parameter labeling points on the vortex string.
The circulation vector ${\vec \Gamma}_s = N_v\kappa {\vec X}'$ of the vortex
is, for positive $N_V$, pointing along the $z$-direction in a
right-handed system.

\begin{figure}[hbt]
\psfrag{sigma}{$\sigma$}
\psfrag{tau}{$\tau$}
\psfrag{X'}{$X'^\mu$}
\psfrag{dotX}{$\dot X^\mu$}
\psfrag{Xmu}{$X^\mu (\tau,\sigma )$}
\psfrag{Spacetime}{Spacetime}
\psfrag{x0}{$x^0$}
\psfrag{x1}{$x^1$}
\psfrag{x2}{$x^2$}
\begin{center}
\includegraphics[width=0.8\textwidth]{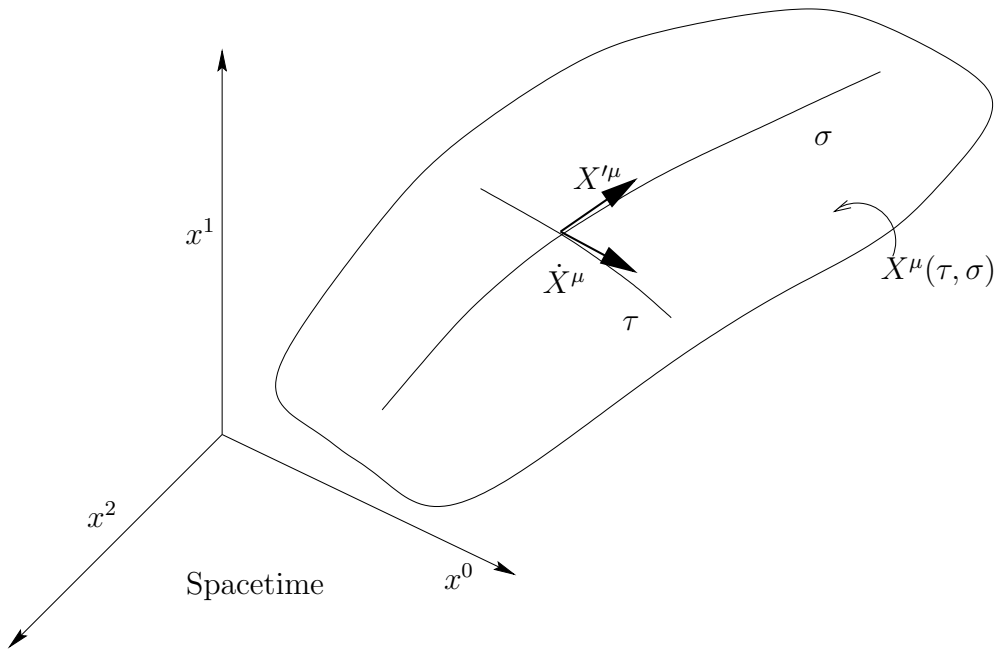}
\end{center}
\caption{\label{embed} The vortex is represented by the string world
sheet embedded in spacetime and hence described by the co-ordinates 
$X^\mu(\tau,\sigma)$. The tangent space basis vectors have
components $\dot X^\mu$ in the timelike and $X'^\mu$ in the spacelike
direction on the world sheet, where the metric is given
by $\gamma_{ab}=
g_{\mu\nu}\,{\partial X^\mu}/{\partial \zeta^a}\,
{\partial X^\nu}/{\partial \zeta^b}$,
with $a,b=1,2$, $\zeta^1=\tau$, $\zeta^2=\sigma$. 
}
\end{figure}

In the second line, the force is written in a form 
manifestly independent of a `mass' value.  
A relativistic generalization of the above expression 
then reads \cite{vilshell}:
\begin{equation}
\left({\bm F}_M\right)_\alpha 
= \gamma_s\, H_{\alpha \mu\nu} \dot X^\mu X'^\nu\,. \label{MforceRel}
\end{equation}
We introduced the relativistic string co-ordinates
$X^\mu=X^\mu(\tau,\sigma)\,,$
and let $X'^\mu \equiv \partial X^\mu/\partial \sigma$ be the line tangent,
as well as $\dot X^\mu \equiv \partial X^\mu/\partial \tau$ the vortex
velocity. 
The essential features of the vortex as a two-dimensional object,
living in spacetime, are represented in Figure \ref{embed}. 

It is to be stressed that both of the forces (\ref{magnusforce}) and
its relativistic counterpart (\ref{MforceRel}) are forces per unit 
$\sigma$-length and are of topological origin. They do not depend on
the local shape of a line segment, represented in the relativistic case by 
the world sheet metric $\gamma_{ab}$, but only on the local external 
field $H_{\alpha\mu\nu}$, generated by other line segments and an
externally imposed flow field (cf. the equation for the corresponding 
action (\ref{Magnusaction}) and the discussion at the beginning of 
section \ref{subsectionLagrangian}). 

The Magnus force 
can be derived from the variation of the action 
(cf., in particular, the Refs. \cite{vilshell,davis2,gradwohl,ben3})
\begin{eqnarray}
S_M&=&\gamma_s \int\! d\tau d\sigma
\, b_{\mu\nu} \dot X^\mu X'^\nu\nonumber\\
   &=& \gamma_s \int\! d^4 x\, b_{\mu\nu} \omega^{\mu\nu} 
\,,\label{Magnusaction}\end{eqnarray}
that is to say
\begin{equation}
\frac{\delta  S_M}{\delta X^\alpha} = 
\gamma_s H_{\alpha \mu\nu} \dot X^\mu X'^\nu
=\left({\bm F}_M\right)_\alpha \,.
\end{equation}
The singular vorticity tensor components are thence given by 
\begin{equation}
\omega^{\mu\nu} \label{omegamunu}
=\gamma_s  \int\!\!\int d\tau d\sigma
\left({\dot X}^\mu {X'}^\nu-{\dot X}^\nu {X'}^\mu\right)
\delta^{(4)}( x-X(\tau, \sigma))\, .
\end{equation}
From the vorticity tensor,
we obtain the components of the usual vorticity vector
by choosing $X^0=t$, 
a choice possible in any global Lorentz frame \cite{ben3}:
\begin{eqnarray}
\omega^{0i} =
\gamma_s \int
d\sigma {X'}^i \delta^{(3)}( \vec x - \vec X(t,\sigma ))\,.
\end{eqnarray}
For a rectilinear line in $z$-direction, $\omega^{0z}=\gamma_s
\delta (x-X) \delta (y-Y)$, which integrated over the $x$-$y$ plane 
gives the circulation. The vorticity
tensor components (\ref{omegamunu}) are, then, to be understood 
as a generalization of a quantity
surface density of circulation, in the sense that 
$\gamma_s = \oint {\bm p}= \int\!\!\int {\bm d}\wedge {\bm p} = 
- \int\!\!\int {^*\!{\bm\omega}}=\int\!\!\int 
-\epsilon_{|\mu\nu||\alpha\beta|}\,\omega^{\mu\nu}{\bm d} x^\alpha 
\wedge {\bm d} x^\beta 
\equiv \int\!\!\int\omega^{\mu\nu} d^2 S_{|\mu\nu|}$. 
Integration is over a (sufficiently small, {\it i.e.} local) 
2-surface with element $d^2 S_{\mu\nu}=
-\epsilon_{\mu\nu|\alpha\beta|}\,{\bm d} x^\alpha 
\wedge {\bm d} x^\beta$, which is threaded by the vortex world sheet.  

\subsection{Electromagnetism of vortex strings}\label{emagnetstrings}
It proves very suggestive to cast the laws we found into the language 
of Maxwell's electrodynamics.
From the familiar Lorentz force law 
\begin{equation}\label{FLorentz}
({\bm F}_{\rm Lorentz})_\alpha=q F_{\alpha\mu}\dot X^\mu\,,
\end{equation}
one can define that 
the `electric' and `magnetic' fields 
are represented by a projection of $H_{\mu\nu\alpha}$ 
on the local vortex axis, {\it i.e.} on the tangent space 
vector along the spacelike direction on the world sheet \cite{myself}:
\begin{eqnarray}
F_{\mu\nu}\equiv H_{\mu\nu\beta} X'^\beta=
\epsilon_{\mu\nu\alpha\beta}j^\alpha X'^\beta\,, 
\nonumber\\
\vec E = 
\rho_0 {\vec u}  \times \vec X' \; ,\quad  
\vec B =  
\epsilon_{0123}\rho_0 u^0 \vec X' 
\,. \label{Fdef}
\end{eqnarray}
The second line is valid for a global Lorentz frame,
in which, as already mentioned above, we can choose $X^0=t$. 
In general, the density $\rho$ 
in the field $H_{\mu\nu\beta}$, and thus also $F_{\mu\nu}$, is arbitrary; 
at the location of the line 
itself and for the Lorentz/Magnus force law,  
however, we set $\rho=\rho_0$. 
In our sign convention for $\epsilon_{0123}=-1$,   
the local `magnetic field' is antiparallel to the local tangent.

The local `electromagnetic' 3-potential 
will accordingly be defined {\it via}
\begin{equation}\label{adef}
a_\mu=b_{\mu\sigma}\equiv 
b_{\mu\nu} X'^\nu
\, .
\end{equation}

In an arrangement of cylindrical symmetry, fulfilled by a ring vortex,
the 3-potential is the vector $a^\mu = (a^0,a^r,a^z)
=(-b_{0\phi},b_{r\phi},b_{z\phi})$.
The component $\psi_S\equiv b_{0\phi}/\rho_0$ corresponds for 
stationary flows and $\rho=\rho_0$ to Stokes' stream function 
\cite{milne-t}. 
The function $\psi_S$  fulfills   
$2\pi \psi_S = 2\pi \int  r (u^r(r,z) dz 
- u^z(r,z) dr)$, according to (\ref{bstreamf}).
The static scalar potential of our `vortodynamics' is thus in the
cylindrically symmetric case given by 
$a^0 = \rho_0\int r(u^z\, dr - u^r\, dz) $.
Integrating 
this relation to obtain the value of $a^0(r_0,z_0)$ at some point
$r_0,z_0$ in the superfluid, the line integrals run over the
envelope of an arbitrary 
surface of revolution, which is 
generated by rotating the integration
line, joining the point at $r_0,z_0$ 
and a point on the $z$-axis of symmetry \cite{milne-t}.

In making the above definitions (\ref{Fdef}), we 
identified $q\equiv\gamma_s = N_v h$ with the `charge'
per unit length of vortex. 
This vortex `charge' is thus of topological nature, quantized
by the index of the homotopy group member (in units with $\hbar=1$,
the charge $q=2\pi N_v$). The quantization of the charge vanishes in 
the purely classical limit of $h\rightarrow 0$, whereas the `electromagnetic'
field strength $\bm F$ 
is understood to be derived from a (superfluid or not) 
conserved number current. 

We can define the vortex current four-vector 
in a fashion analogous to the definition of the potential in (\ref{adef}),  
from the minimal  coupling term represented in (\ref{Magnusaction}):
\begin{equation}\label{jdef}
\Omega^\mu\equiv \omega^{\mu\sigma} \equiv
\omega^{\mu\nu} X'_\nu\,,
\end{equation}
that is, we project the vorticity current on the local vortex axis.  
This completes the picture of the electromagnetism analogy we wanted to
develop. 
We have constructed, within the
semiclassical realm, a fundamental 
topological object, the vortex, which moves according to familiar
laws of electrodynamics, by projecting 
the string equations of motion on the local spacelike vortex tangent axis. 
The field equations have the form 
of Maxwell equations local in $\sigma$,
\begin{eqnarray}\label{Maxwelleq}
{\bm d}\wedge {\bm F}=0\;, \qquad ^*\!{\bm d}\wedge {({\hbar^2}
{^*\!{\bm F}}/K)}
= {\bm \Omega}\,,
\end{eqnarray}   
where the field ${\bm F}$ 
is the usual 2-form and the current 
${\bm \Omega}$ 
a 1-form, whose components are defined 
in (\ref{Fdef}) and (\ref{jdef}), respectively. 
The vortices can be understood as fundamental, `charged'  objects, 
generating `electromagnetic' fields, 
which in turn act on them by the local Lorentz force in 
(\ref{FLorentz}).\footnote{We remark that we have chosen not to 
incorporate the factor $4\pi$, appearing in the cgs units 
Maxwell equations, into the definition of the vortex charge, 
{\it i.e.} wrote the 
`Maxwell' equations 
in Heaviside-Lorentz units \cite{schmutzer}. } 

The equations (\ref{Maxwelleq}) are 
{\em representations} of the conservation laws of particle number, vorticity,  
and the equation (\ref{pmudef}) characterizing properties of the medium, 
which relates these conservation equations to each other. The homogeneous 
`Maxwell' equation is a projection of number conservation, 
whereas the inhomogeneous `Maxwell' equation is equivalent
to a projection of (the conservation of) vorticity, 
combined with a property of the medium.
{\it Vice versa}, 
we conclude that the Maxwell equations of ordinary 
electromagnetism can be cast into the form of conservation 
equations of relativistic perfect fluid hydrodynamics, provided
we admit the identifications specified in (\ref{dualtrans}).  

\subsection{Comparison of relativistic and nonrelativistic notions}
The identifications we have made pertain to the vortex as a stringlike
fundamental object and are suitable for a relativistic framework. In
the remainder of this paper, however, we will only be concerned with
nonrelativistic Eulerian flows.\footnote{The term `Eulerian' 
is used for a fluid 
obeying in its (non-dissipative) dynamics a continuum version 
of Newton's equation of motion, the Euler equation 
(there exist 
also generalizations of this equation for curved 
space-time backgrounds, cf. \S 22.3 in \cite{MTW}).}   
At this stage it is thus useful to compare and contrast the 
relativistic notions and those of classical nonrelativistic 
hydrodynamics \cite{lamb,milne-t}. 

The most important difference of the two frameworks occurs if we
consider the two definitions (\ref{Gammas}) and (\ref{gammas}) for the 
circulation. The {\em kinematical} quantity {velocity} suitable for its
definition in nonrelativistic circumstances can not be used for 
its relativistic definition. 
The {\em dynamical} quantity average angular momentum per particle, though,  
which is quantized into units of $\hbar$ because of the existence of a
macroscopic superfluid phase, is a well-defined quantity for any 
superfluid. Because of this fact, the definition in (\ref{gammas}) 
is the truly invariant definition of circulation. The only possibility 
to use a `velocity line integral' for the definition of circulation
is to employ the perfect fluid relation $p_\alpha = \mu u_\alpha$. 
Then, however, one has the
complication of a possible position dependence of $\mu$. 

A prescription to translate from the fully relativistic 
superfluid to terms of standard nonrelativistic 
hydrodynamics is afforded by the limits 
$\mu\rightarrow m$, $u^\alpha\rightarrow (1,v^i_s)$, 
and the replacement $\gamma_s\rightarrow \Gamma_s$.  
\section{Canonical structure of vortex motion}\label{canonstructure}    
\subsection{Phonons and the Vortex Mass}
In quantum electrodynamics, the fundamental entities interact between each
other and with themselves by photons. The corresponding quanta in our 
view of a superfluid are phonons, the excitations 
of the `vacuum ether' surrounding the `charged' strings. The  role of 
these excitations  is addressed in this section.   

The vortices acquire a nonzero
hydrodynamic mass from their self-inter\-action with these phonon excitations.
This is most easily understood if one realizes that in a  
2+1d superfluid there exists a direct correspondence between the genuinely
relativistic electron-positron pair interaction {\it via} photons in quantum
electrodynamics and a `relativistic'
point-vortex-point-antivortex
pair interaction {\it via} 
phonons \cite{popov2,arovas} (we will use throughout
inverted commas to distinguish 
pseudorelativistic $\equiv $ `relativistic' behaviour from the 
actual Lorentz invariance of section \ref{secdual}). 
In the light of the vortex $-$ fundamental
object correspondence expounded in the preceding section this is only
natural: The only `vacuum' excitations
(`vacuum' in the sense of the spontaneously symmetry-broken 
superfluid vacuum), by means of which different vortex line
segments can interact are, in the hydrodynamic limit, phonons. What
remained to be done is to convert the quite direct 
2+1d correspondence into a 3+1d
correspondence by defining the respective quantities local on the
string, {\it i.e.} as functions of $\sigma$. Then, the
phonons give `relativistic' fields propagating on a nonrelativistic 
background of Eulerian superflow, and mediating interactions between 
(singular) vortex line segments. 

To explain this further, we begin by considering that  
the unit circulation 
vortex carries, per unit length, the hydrodynamic self-energy 
\begin{equation}\label{Energy}
E_{\rm self}=  
\frac{h^2\rho_0}{4\pi m}
\left[\ln \left(\frac{8R_c}{\xi e^C}\right)
\right]\,
\end{equation}
with it. The energy $E_{\rm self}$ 
is the energy of a vortex sitting at some
fixed place, that is, its rest energy in its rest frame.

The infrared cutoff in the logarithm is in the static limit 
equal to the mean distance of line elements, 
respectively, in the localized self induction approximation, 
proportional to the 
local curvature radius $R_c$ of the line. 
The constant $C$ parameterizes the core structure \cite{robgrant}, 
and has order unity. In classical hydrodynamics,
with a hollow core of radius $\xi$, $C=2$, whereas in a model of constant
core vorticity one has $C=7/4$. In the Gross-Pitaevski\v{\i} framework 
(\cite{pita},\cite{gross3}), 
for singular vorticity, it turns out that $C=1.615$ \cite{robgrant}. 

If we separate off in the order parameter phase two parts, 
\begin{equation}
\theta = \theta_{bg}+\theta_{ph}\,, 
\end{equation}
a part due to a background flow $\theta_{bg}$, and a part 
which describes sound excitations $\theta_{ph}\ll
\theta_{bg}$ on this background, 
we can carry through the program of the last sections for 
$\theta_{bg}$ and $\theta_{ph}$ separately. 
For the phase $\theta_{bg}$, we can not retain the Lorentz invariance of
the equations we derived there (our fluid is very much Eulerian),  
and have to use the Galilei invariant set of equations 
(\ref{bijk})-(\ref{stream}). 
On the other hand, for the part $\theta_{ph}$, a `relativistic' 
equation from which to start the dual transformation 
derivation for $\theta_{ph}$
is the wave equation 
\begin{equation}\label{eqmotioncs}
\partial_\mu\partial^\mu\theta_{ph}\equiv 
\left[-\frac{1}{c_s^2}\frac{\partial^2}{\partial t^2}
+\Delta\right]\theta_{ph}=0 \,.
\end{equation}
This is `relativistic' in the sense that the speed of light $c$
is replaced by the speed of sound $c_s$ in a Lorentz invariant 
scalar wave equation for $\theta_{ph}$.
From this part of the 
phase we can derive `electromagnetic' phonon wave field strengths obeying
the `relativistic' Maxwell equations \cite{davis2}. 

The nonlinear equation of motion of the superfluid,
the Euler equation, gives in its linearized version, together
with the continuity equation, the above equation of motion 
(\ref{eqmotioncs}) for $\theta_{\rm ph}\ll \theta_{bg}$, if
the background flow is at a velocity much less than that of sound. 
For general background flows 
with velocities $(\hbar/m)|\nabla\theta_{bg}|\lesssim c_s$, there results 
a scalar wave equation for sound 
in a curved Lorentzian
signature background metric with nonzero curvature. 
See the nice discussion in \cite{visser2} (also cf. section IV in 
\cite{langlois1}).  
We consider, however, `nonrelativistic' 
background flows of small Mach number, 
so that these corrections are not of importance here (though they may be
important in other contexts like for the Aharonov-Bohm interferences of phonons
leading to the Iordanski\v{\i} force \cite{grishaiord}), and the 
`acoustic metric' is taken to be that of a global Lorentz frame. 

We now have defined the entities vortices as objects 
defined by the superfluid vacuum
background and `relativistically' interacting by the small density and
phase perturbations in this vacuum medium. 
Hence, they obey the Einsteinian mass-energy relation 
\begin{equation}\label{M0}
M_0 c_s^2= E_{\rm self}\,.
\end{equation}
The mass $M_0$ is the hydrodynamic vortex mass in the rest frame of
a vortex in equilibrium. 
If we consider a neutral, unpaired superfluid, this mass is dominant
compared to other possible sources as the so-called
backflow mass or the core mass
corresponding to the normal fluid in the core \cite{grishabaym}. This is 
true because we are in the limit of $\xi \ll R_c$,  
in which the logarithmic  divergence of the hydrodynamic mass
dominates other possible contributions. In field theoretical terms, 
the energy contribution of the 
Goldstone boson field ({\it i.e.} the phonon field) is not 
screened by the gauge field like in superfluids with a local dynamical
gauge field, which are charged in a conventional sense.   
This leads to the logarithmic divergence of the
energy associated with the spontaneously broken global symmetry. 

It is useful to compare the definition of the vortex mass above
to that of the classical electron radius $r_c$ in electromagnetism. 
For a homogeneously charged electron of radius $r_c$, this definition
reads $m_e c^2 = (3/5) e^2/r_c$, where the factor 3/5 stems from 
the electron (`core') structure, which was assumed to be homogeneous.  
In the case of the electron, the mass serves to define, {\it via}  
the expression for the field energy, the value of the classical
electron radius. In the case of the vortex (a line `charge'), 
the field energy of
flow around the vortex serves to define the renormalized,
hydrodynamic vortex mass and a cutoff
needs to be introduced because this field energy is logarithmically
divergent. 

The knowledge of the mass $M_0$ will be sufficient for our purposes,  
as the formalism laid down in these
pages should apply for the actual, dense superfluid and consequently 
is restricted to velocities of the vortex and the background
superfluid much less than the speed of sound, indeed even much less
than the Landau critical velocity of roton creation. Also,
we will consider scales much larger than $\xi$. 
Correspondingly, the (local) frequencies of vortex motion will be much less
than the typical frequency $\omega_s \equiv c_s/\xi$. 
Scaling this frequency with parameters appropriate for helium II gives
\begin{equation}\label{omegasfirst}
\omega_s=
\frac{c_s}{\xi}= 9.4\cdot 10^{11} {\rm Hz}\, \frac{c_s[ 240\, {\rm m/s}]}
{\xi[\sigma_{LJ}]}\,,
\end{equation}
which turns out be 
very close to the roton minimum ($\omega_{\rm roton}\simeq 1.13\cdot
10^{12}\,$Hz at $p\simeq\,$1 bar). We have scaled the coherence length 
with the Lennard Jones parameter of the helium interaction, $\sigma_{LJ}=$
2.556 \AA, and assume ${\xi[\sigma_{LJ}]}$ is approximately unity.  
The condition of `nonrelativistic' 
velocities on scales much larger than $\xi$ is then 
equivalent to $\omega\ll \omega_s$.
For `relativistic' vortex motion frequencies 
in the order of and above $\omega_s$, 
the vortex mass will be frequency and wave vector dependent 
\cite{arovas}.
These high frequencies, 
employed in a semiclassical treatment, may be of
interest in dilute superfluids.  
In what follows, we however restrict ourselves to describe 
`nonrelativistic' vortex motion, which is the case of interest 
in a dense superfluid. 

In the context of the question of vortex mass, it is advisable
to point out that it is unnecessary, and indeed misleading, to use 
any {postulated} dynamical behaviour for the order parameter 
in the Lagrangian form (for example, of the Gross-Pitaevski\v{\i} 
or time dependent Ginzburg-Landau variety).  
If we discuss the proper way of deriving the mass (\ref{M0}) 
in a dense superfluid, we just need 
that the phase moves according to the Josephson relation 
$\hbar\dot\theta = -\left(\mu + \frac12 m{\vec v}_s^2\right)$
\cite{duan}.  
The validity of the Josephson relation, in turn, is not dependent 
on a Lagrangian for the order parameter, and is derivable from (Hamiltonian) 
superfluid hydrodynamics alone \cite{khalatni}.

\subsection{The Vortex Lagrangian}\label{subsectionLagrangian}
We have argued that the vortex has mass.
Being a line-like, stable topological object, we assume it further to
develop restoring forces if it 
is deformed, starting from some equilibrium position. That is, the vortex 
is assumed to have elastic energy, arising from the (local)
interaction of a particular line segment with adjacent line segments. 

Let us first spend a few words on the relation of the self-action 
we will write down to `relativistic' vortex motion properties. 
In the `nonrelativistic' case under study, the differential
string arc length will be written as $\sqrt{\gamma
(t,\sigma)}d\sigma $ (we remind the reader that the inverted commas are used
to distinguish pseudorelativistic from proper relativistic terms).
The self-action 
of the singular Nambu string\footnote{We refer to this
elementary extended object for brevity simply as `Nambu' string, 
though it has been introduced independently by Nambu and Got{\=o}
\cite{goto,nambustrings}, and thus is (properly)  also quite often 
referred to as `Nambu-Got\=o' string.} 
is proportional
to the world sheet area \cite{vilenkin1}:
\begin{equation}\label{Nambu}
S_{\rm Nambu} = - \mu_T\int\!\!\int \sqrt {-\gamma(\zeta^0,\zeta^1)}\, 
d^2\zeta\,, 
\end{equation}
which is given by the double integral of
the square root of the negative world sheet metric determinant 
$\sqrt{-\gamma}$ over the timelike $\zeta^0=\tau$ and spacelike 
$\zeta^1=\sigma$ co-ordinates, parameterizing the world sheet
(cf. Figure \ref{embed}). 
The Nambu action is minus the constant 
string tension $\mu_T$ ({\it i.e.}, a constant
self energy per unit length),
multiplied by this area \cite{vilshell,vilenkin1,teitelboim}. 
The writing for the string arc length we will employ thus indicates
that we are dealing with a `Galilean 
limit' of the `world sheet area' interval, in which the
only metric element is the proper length interval of the line. 

The Nambu string is structureless ({\it i.e.}, the core extension is 
negligibly small on the scales of interest), and has constant string
tension. The actual superfluid vortex has a
(quantum mechanical) core structure of extension $\xi$ which is comparatively 
large\footnote{Whereby it is meant that 
the characteristic velocity at the core circumference,  
$v_L = \kappa/2\pi \xi$, is much less than the speed of sound, if we again 
take $\xi \sim \sigma_{LJ}$.} 
and, in addition, has a string tension (the vortex mass) that 
depends on vortex co-ordinates through the cutoff in the
renormalization logarithm. 
Then, on the large ($\gg \xi$) scales we have to consider the vortex can
not be considered as a `relativistic' Nambu string 
living in the spontaneously broken symmetry vacuum
of the dense superfluid. 
Given that the velocities the vortex reaches are on
these scales always less than $c_s$, as a result,  
the velocity dependent 
part of the vortex self-action is dominated by that 
of the Magnus action. This corresponds to the dominance of the minimal 
coupling term in the momentum over the kinetic part, see equation 
(\ref{ratioP}) below. 

\subsubsection{Co-ordinate frames}
The co-ordinates we will be  using to describe the vortex
self-action are defined by local right-handed basis vectors on the
string. We can, for example, 
choose them to be the triad
\begin{equation}\label{triad}
{\vec e}_1 (\sigma) = {\vec e}_2 \times {\vec \tau}\,,\;\;\;
{\vec e}_2 (\sigma) = -\frac 1\gamma\, {\vec X}''\,,\;\;\;
\vec\tau = \frac 1{\sqrt\gamma}\,{\vec X}'\,, 
\end{equation}
that is, the binormal, negative normal 
and tangent unit vectors of the line. 
They are related by the Serret-Frenet formulas \cite{aris}
to fundamental 
invariant properties of the line
\begin{equation}
\frac 1{\sqrt\gamma}\,{\vec \tau}\,'= -\frac 1{R_c} \, {\vec e}_2\,,\;\;\;
\frac 1{\sqrt\gamma}\,{\vec e}_2\,'
=\frac 1 R_c \, {\vec \tau}- T{\vec e}_1\,,\;\;\;
\frac 1{\sqrt\gamma}\,{\vec e}_1\,'= T{\vec e}_2
\,,
\end{equation}
where $R^{-1}_c (\sigma)=|{\vec X}''/\gamma|$ is the curvature ($R_c$ the
curvature radius), and $T(\sigma)$ the torsion of the line.  
For a three-dimensional superfluid a typical equilibrium 
line configuration is a circular vortex, for which 
${\vec e}_1={\vec e}_Z,\,{\vec e}_2={\vec e}_R,{\vec \tau}
={\vec e}_\Phi$,
in a conventional cylindrical co-ordinate system
with the normalised triad above. 

If it appears more convenient
for representation purposes, we will use a more general system of 
non-normalised co-ordinate basis vectors, which are defined formally by 
the derivatives \cite{MTW} 
\begin{equation}\label{coordbasis}
{\vec e}_{\it a}=\partial/\partial X^{\it a}\,,
\quad {\vec e}_\sigma=\partial/\partial \sigma\;\qquad ({\it a}={\it
1,2})\,,
\end{equation}
in our non-curved Euclidean embedding 
space simply acting on the position vector
$\vec X$, thereby 
creating the basis. 
We will freely use the more convenient system, indicating the type
of system by indices $a=1,2$ for the first normalised triad and 
$\it a =1,2$ for the latter general co-ordinate basis.   

\begin{figure}[hbt]
\psfrag{sigma}{$\sigma$}
\psfrag{sigma0}{$\sigma_0$}
\psfrag{kappa}{$\kappa$}
\psfrag{Q}{${\vec Q}(\sigma_0)$}  
\psfrag{Q2}{$Q^{\it 1}(\sigma_0)$}
\psfrag{Q1}{$Q^{\it 2}(\sigma_ 0)$}
\psfrag{X'}{${\vec X}' (\sigma_0)$}
\psfrag{e2}{${{\vec e}_{\it 1}}(\sigma_0)$}
\psfrag{e1}{${{\vec e}_{\it 2}}(\sigma_0)$}
{\includegraphics[width=\textwidth]{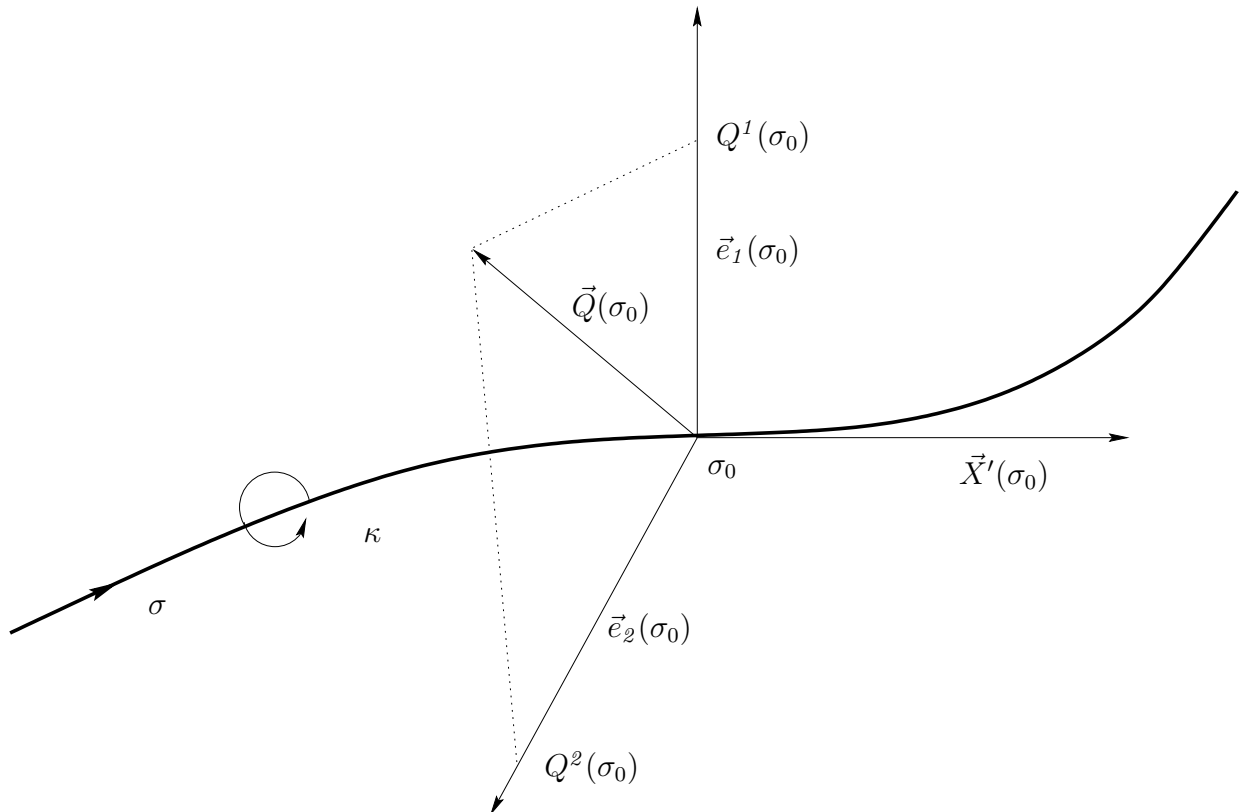}}
\caption{\label{cosys}
Co-ordinate ortho-basis 
on the vortex line. The displacement
vectors $\vec Q(\sigma_ 0)$ are lying in the plane spanned by 
${\vec e}_{\it 1}(\sigma_ 0), {\vec e}_{\it 2}(\sigma_ 0)$,
perpendicular to the local tangent ${\vec X}'(\sigma_0)$. Normalising 
these vectors, one obtains the basis (\ref{triad}).}
\end{figure}
\subsubsection{The self-action}
After this preparatory work of fixing conventions, we write down 
the vortex self-action as a sum of its static part and one 
quadratic in derivatives of perturbations from the equilibrium string
configuration:
\begin{equation}\label{self}
S_{\rm self}[\vec Q(t,\sigma)]=
-\int\!\oint dt d\sigma\sqrt{\gamma(t,\sigma)}\left\{ M_0 c_s^2 
-\frac12 M_0 \dot{\vec Q}{}^2 +\frac12\frac\alpha{\gamma}\, 
{\vec Q}'{}^2\right\}
\,.
\end{equation}
The displacements ${\vec Q}$ perpendicular to the line 
have to be small. 
Only in this case of small perturbations from an    
equilibrium configuration the equations of motion of interacting  
vortex segments obey the Hamiltonian structure we wish
to derive below \cite{fettquth}. In
particular, self-crossings of a line bending back on itself
have to be excluded. 
The first terms in the self action then represent 
`nonrelativistic' terms of static and kinetic energy in the Lagrangian
for a particle, weighted with the local differential string arc length, and 
integrated over the length of the string. 

The cutoff related elastic coefficient 
$\alpha$ (parameterized by the longitudinal  
as well as the transversal string core structure), 
depends on $t,\sigma$ in general, just as the vortex effective
hydrodynamic mass does. In accordance with our `nonrelativistic' treatment,  
we do not consider short wavelength perturbations, 
where this dependence becomes significant, and take 
$\alpha\equiv M_0 c_s^2=E_{\rm self}$. 
This specific choice in the elastic energy of the string  
is related to a cutoff choice in the 
localized self induction approximation of classical
hydrodynamics \cite{schwarz1}. It corresponds to the assumption
that in the long wavelength 
limit we are using, only massless
sound excitations propagating along the string can
survive\footnote{This is true given that we consider the kinetic 
and elastic parts isolated from the static energy. 
Neglecting the kinetic energy, we can obtain the Kelvin modes \cite{fettquth}. 
Apart from the fact that the vortex has a certain core structure 
(such that an ultraviolet cutoff is present), they are
not related to compressibility, {\it i.e.} they exist even for infinite
$c_s$.} \cite{foerster}.  
It also naturally accounts for the fact that $\alpha$ 
must remain finite in the incompressible limit 
$c_s\rightarrow\infty$ (whereas the hydrodynamic mass has to vanish).

For a simple straight line in $z$-direction, we can write for (\ref{self}), 
Fourier analysing the 
co-ordinates $\vec Q = (1/(2\pi)^2)\int d\omega\int dk 
\exp[-i(\omega t -kz)]\vec Q (\omega ,k)$, and neglecting the
dependence of $M_0$ on the co-ordinates,  
\begin{equation}\label{selfomegak}
S_{\rm self}/ {M_0 c_s^2}= -\int\!\int dt\, dz 
+\frac1{(2\pi)^2}\int\!\int\! d\omega\, dk\,\frac12 {{\vec Q}^2(\omega ,k)}
\left\{ (\omega/c_s)^2 - k^2 \right\}
\,.
\end{equation}
If the wiggles of the line are occuring on scales $\ll R_c$, this 
should also hold for other shapes of line by summing up contributions
of  approximately straight segments. 
\subsubsection{The total vortex action and momentum} 
To the self-action, we have to add the interaction with the
background. Using the conventions (\ref{psiidef}),(\ref{adef}),
\begin{eqnarray}
S_M & = &  \label{SMpsi}
\gamma_s \int\!\!\oint dt\, d\sigma 
\left(\rho_0\psi_i X'^i + b_{ij}X'^j \dot X^i \right)\\ 
& \equiv & q\int\!\!\oint dt\, d\sigma 
\left(-a^0  + a_i \dot X^i \right).\nonumber
\end{eqnarray}
The gauge potential $(\psi_i,b_{ij})$ 
is defined to belong entirely to the background:
\begin{equation}\label{background}
b_{\mu\nu}(x)\equiv \frac1{(2\pi)^4}\int_{{\bm k}<{\bm k}_0} \!\!d^4 k\, 
\tilde b_{\mu\nu}(k) \exp [ik_\mu x^\mu]\,. 
\end{equation}
The cutoff $(k_\mu)_0=(-\omega/c_s,k_i)_0$ 
indicates the separation line 
between what we consider as being a phonon, which is integrated out in the
self-action, and what we lump together
into a time dependent and inhomogeneous background.
In what follows, we will neglect the remaining phonon fluctuations 
and approximate the background to be an incompressible superfluid. The 
separation line is fixed in a quite natural way 
by the infrared cutoff in the vortex 
energy (\ref{Energy}), $|\vec k_0|=\exp C/8 R_c$.
 
Summation of (\ref{SMpsi}) and (\ref{self}) yields the total vortex action 
\begin{equation}\label{SV}
S_V=S_{\rm self}+S_M\,.
\end{equation} 
Taking the functional derivative of $S_V$ after the vortex velocity, we
arrive at the following expression for the vortex momentum per
$\sigma $-length interval of the vortex line 
($\dot{\vec X} = \dot {\vec Q}$): 
\begin{equation}\label{Pcanon}
\vec P = \vec P^{\rm inc} + \vec P^{\rm kin}
=q \vec a + M_0 \sqrt\gamma \dot {\vec X}
\, .
\end{equation}
It has, as expected, the same appearance as the canonical momentum 
for a nonrelativistically moving particle of charge $q$, which is 
subject to an external vector 
potential ${\vec a}$. We separated the momentum into a 
part due to the (incompressible) background $\vec P^{\rm inc}$ 
and a kinetic (vortex matter) part $\vec P^{\rm kin}$. 

Corresponding to the dominance of the static energy in the
self-action, the ratio of the contributions to the momentum 
is in order of magnitude 
\begin{equation}
\frac{\,\left|{\vec P}^{\rm kin}\right|\,}
{\,\left|{\vec P}^{\rm inc}\right|\,}
\approx \frac{\Gamma_s}{|{\vec X}|c_s}\frac{|\dot {\vec X}|}{c_s}\,.
\label{ratioP}
\end{equation} 
This depends on 
$|\dot {\vec X}|/c_s$ as well as $\kappa/(c_s|{\vec X}|) $ 
($= {O}(\xi/|{\vec X}|)$ in helium II). 
Both quantities are necessarily $\ll 1$ if the vortex is to be
described within the hydrodynamic formalism we presented. 
We used here that $|b_{ij}|$ is of 
order $|X^k|$ in a Cartesian frame (see below), and 
neglected the dependence of the self-energy 
logarithm on the vortex co-ordinates (which multiplies the right-hand
side of the equation above). 
\subsection{Gauge dependence of the vortex momentum}\label{Gaugedepend}
The momentum (\ref{Pcanon}) is gauge dependent 
through the choice for the vector potential $\vec a$. The relation (\ref{bijk})
determines $\vec a$ in terms of the original field $b_{ij}$. 
This equation is, in turn, equivalent 
to the last expression in (\ref{Fdef}) in its nonrelativistic 
version, 
\begin{equation}\label{rotaBX'}
{\rm rot}\,\vec a =\vec B=-\rho_0\vec X '\,.\label{rotaB}
\end{equation}
 
In a Cartesian frame, a possible solution for $b_{ij}$ is 
$b_{ij}=-(1/3)\rho_0 n_{ijk}X^k$ (it is, for example, used 
to represent to vortex velocity dependent part of the Magnus action
in \cite{rasetti}). 
This is an isotropic solution,
which is convenient to describe the purely spatial part of the 
gauge 2-form in the bulk superfluid. 
In the presence of the vortex, however, we choose another solution more 
appropriate to a natural basis on the string. 
For a singly quantized vortex, 
the momentum components $P_{\it a}$ in the 
basis of (\ref{coordbasis}), ${\vec e}_{\it a}$, 
${\vec e}_\sigma= {\vec X}'$  (${\it a}={\it 1,2}$), with determinant
$g={\rm det}[{\vec e}_i\cdot {\vec e}_j]$,    
obey \cite{geo}
\begin{equation}\label{p1p2}
\partial_{\it 2} P^{\rm inc}_{\it 1} -\partial_{\it 1} P^{\rm
inc}_{\it 2}
=h \rho_0\sqrt g\,, 
\end{equation}
as follows from (\ref{rotaB}). 
A factor ${N_v}$ in front of the right hand side
enters for vortices of arbitrary winding number.

An isotropic solution of (\ref{p1p2}) in Cartesian co-ordinates 
reads $\vec P^{\rm inc} = (1/2)h\rho_0 \vec X \times \vec X'$. 
The simplest possible 
solutions are obtained if we gauge one of the $P^{\rm inc}_i$'s to zero:
\begin{eqnarray}
P^{\rm inc}_{\it 2}=0 &\, \Rightarrow\, & P^{\rm inc}_{\it 1}=h\rho_0\int
dX^{\it 2} \sqrt g \;,\nonumber\\
P^{\rm inc}_{\it 1}=0\; &\, \Rightarrow\, &  P^{\rm inc}_{\it 2}
=-h\rho_0\int dX^{\it 1} \sqrt g \,.\label{p1p20}
\end{eqnarray} 

It is stressed that the canonical 
local momentum $\vec P^{\rm inc}$ is a gauge object,
and not necessarily identical with a physical momentum of the vortex.  
A connection to a physical momentum can be established by 
using the first choice in the equation above 
for a circular vortex (${\vec e}_{\it 1}={\vec e}_Z, \, {\vec e}_{\it 2}
={\vec e}_R,\,{\vec e}_\sigma=
\vec X'={\vec e}_\Phi$ and $\sqrt g = r$), and integrating 
$P^{\rm inc}_{\it 1}=h\rho_0\int dX^{\it 2} \sqrt g$ over 
$\Phi=\sigma=[0:2\pi]$.  
We then obtain that $\vec P^{\rm Kelvin} = h\rho_0 \pi R^2 {\vec e}_Z$ 
equals the Kelvin momentum.  For a superfluid
the Kelvin momentum is just the surface area of the vortex times the bulk 
superfluid density times Planck's quantum of action. This result 
is also obtained by integrating  
${\vec P}=\int\!\int\!\int \!m\rho_0{\vec v}_s\,dV=
\hbar \int\!\int\!\int \!\rho_0\nabla \theta\, dV=
\rho_0 \hbar \int\!\int \!\theta\, d{\vec S}\,,$ and taking the 
$2\pi$-discontinuity of $\theta$ (cf. Figure \ref{vortexisolated}) 
across the surface enclosed by the vortex line.

\subsection{Quantization of Vortex Motion}
The canonical quantization of the {\em massive} vortex string proceeds in the
usual manner. We take as the canonical phase space the displacement vector 
$\vec Q$ together with the momentum $\vec P$ from (\ref{Pcanon}). 
If we are to 
quantize the vortex motion, we have to impose the following canonical 
commutation relations, 
written for convenience in the basis (\ref{triad}), 
$$
[Q^a(\sigma),P_b(\sigma ')] =
i\hbar\delta^a{}_{b}\delta(\sigma-\sigma ')\,,\quad 
(a,b=1,2)\vspace*{-0.5em}
$$
\begin{equation}
[Q^a(\sigma),Q^b(\sigma ')] = 0 \label{canoncomm}
\end{equation}
$$
[P_a(\sigma),P_b(\sigma ')]= 0
$$
It should be observed that in our `magnetic' field, the kinetic momentum
components, defined in (\ref{Pcanon}),
do not commute, as follows from the classical Poisson bracket, respectively 
from direct calculation, using the commutators above:
\begin{equation}
[P^{\rm kin}_a(\sigma),P^{\rm kin}_b(\sigma ')]= 
i\hbar q(-\rho_0) n_{ab}\delta(\sigma-\sigma ')\,,
\end{equation}   
which amounts to saying that the local velocity 
components of the string 
are not both determined to arbitrary accuracy. This statement 
is analogous to that for the velocity components 
of an electron in a magnetic field (cf. \cite{ll}, \S 110), quite as it should,
according to the analogy explained in  section \ref{emagnetstrings}.   

  
\begin{figure}[hbt]
\psfrag{a}{\small $\alpha(\sigma)$}
\psfrag{sigma}{$\sigma$}
\psfrag{kappa}{$\kappa$}
\psfrag{Q}{${\vec P}^{\rm inc}$}  
\psfrag{Q1}{$P^{\rm inc}_{\it 1}$}
\psfrag{Q2}{$P^{\rm inc}_{\it 2}$}
\psfrag{X'}{$\vec X'$}
\psfrag{e1}{${{\vec e}_{\it 1}}$}
\psfrag{e2}{${{\vec e}_{\it 2}}$}
{\includegraphics[width=\textwidth]{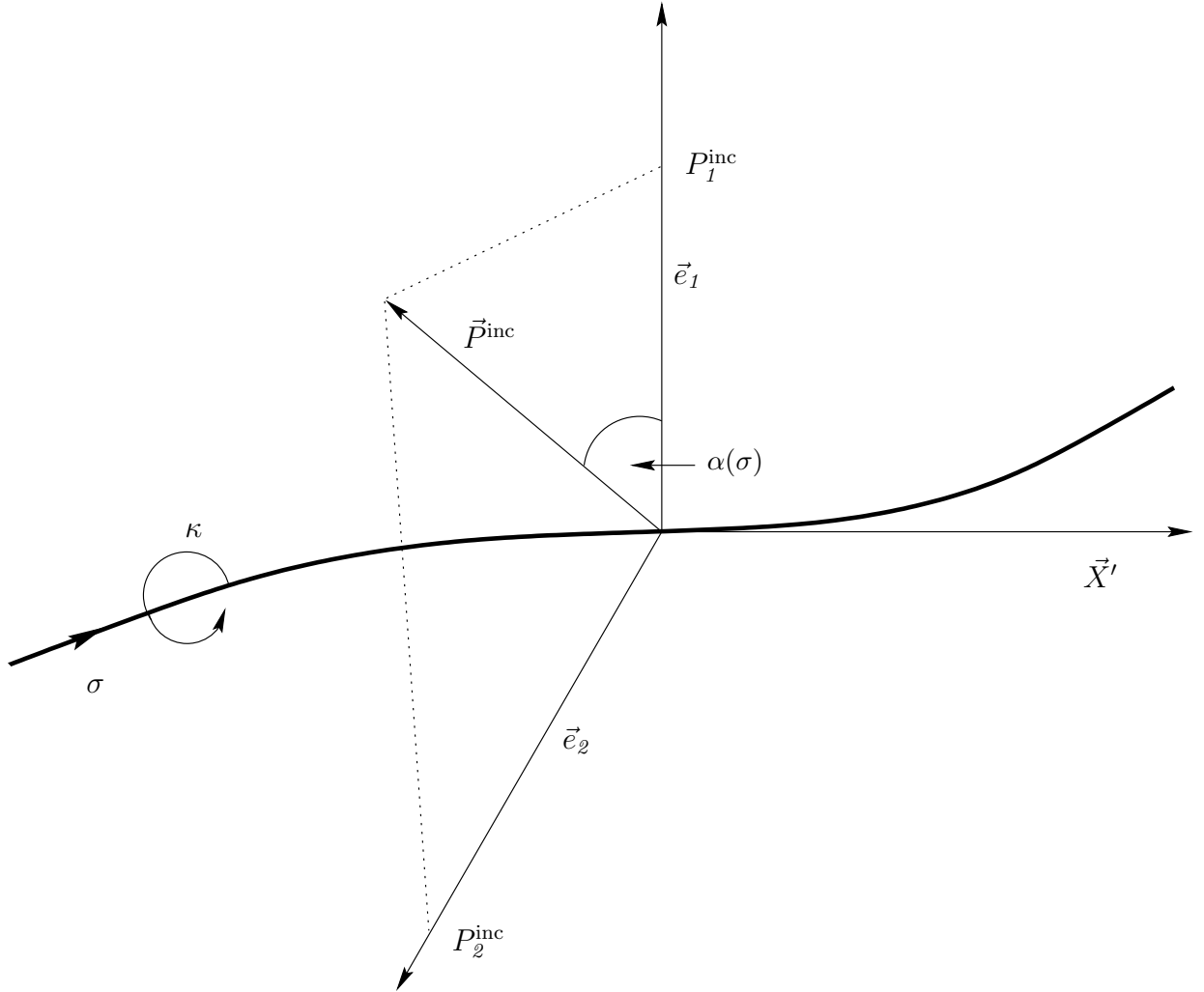}}
\caption{\label{mombasis} The direction of the incompressible part of
the vortex momentum in co-ordinate space depends on the choice of the gauge in 
(\ref{rotaB}) respectively (\ref{p1p2}).    
On every point $\sigma$ of the vortex line the momentum can point in a
different direction of the local co-ordinate plane 
${\vec e}_{\it 1},\, {\vec e}_{\it 2}$.  
This direction is parameterized by the angle $\alpha (\sigma)$.
The first choice in (\ref{p1p20}) corresponds to $\alpha =0 $, the second
to $\alpha =\pi/2$.}
\end{figure} 
If we neglect the 
part $\vec P^{\rm kin}$ as compared to the dominant $\vec P^{\rm inc}$
in (\ref{Pcanon}) altogether, as indicated by (\ref{ratioP}), 
taking again a ring vortex as the simple 
archetypical example, the complete quantum dynamics of the 
(undeformed) vortex is effectively one-dimensional (has only one 
independent co-ordinate and momentum component), 
and given by the canonical commutator 
$$
 [Z (\sigma),P^{\rm inc}_Z(\sigma')] = 
 \frac12 [Z,h\rho_0 R^2]=i\hbar \delta (\sigma -\sigma')
$$\vspace*{-1.5em}
\begin{eqnarray}
[{Z}, {S}] =  i (2\pi \rho_0)^{-1}\,,
\end{eqnarray}
where the second line is the version of the local commutator in the
first line integrated along the string, 
so that the commutator involves the Kelvin momentum of the ring. 
Consequently, $S = \pi R^2$ represents a surface operator of the ring plane.
In the limit of $\vec P^{\rm kin}\rightarrow 0$, generally, phase space and 
configuration space merge and become indistinguishable, the momentum 
components, then, becoming functions of the co-ordinates alone. 

The momentum space counterpart of the Figure \ref{cosys} is depicted in 
Figure \ref{mombasis}. We see that on every point of the line, we are 
free to choose the direction of the momentum afresh, according 
to a solution of (\ref{p1p2}) with some local gauge. 

\subsection{The Hamiltonian and the equations of motion}
\label{Hameqmotion}
The vortex moves without dissipation, that is, there
is no damping force term in the collective co-ordinate equation of 
motion for the vortex. Hence we can write 
down a conserved vortex energy and thus a vortex Hamiltonian. 
In the present case, it is useful to take the Coulomb gauge
div\,$\vec a= {\rm div}\,\vec P^{\rm inc}=0$ 
for the representation of the Hamiltonian: 
%
\begin{eqnarray}
H_V 
& = &  \oint\! d\sigma 
\sqrt\gamma\, \left[M_0 c_s^2  
+ \frac 1{2\gamma M_0} \left( {\vec P}
-q{\vec a}\right)^2\right.\nonumber \\
& & \quad \,\quad \qquad \left. +\frac{M_0 c_s^2}{2\gamma}\,
{\vec Q}'{}^2\right] 
+ q\int\! d\sigma \left(\frac12 a^0_C 
 +a^0_{u}\right)\,.
\label{HV}
\end{eqnarray}
We separated Coulomb and background velocity 
parts of the scalar potential, {\it i.e.} wrote
 $a^0\equiv a_C^0+a^0_{u}$\,.
This gives the correct factor of 1/2 in the Coulomb interaction energy 
with other vortices: 
The energy of these other vortices 
is contained in the background part of the energy,
whereas $(1/2) q a_C^0$ is the energy solely pertaining to the vortex under
consideration.
The Hamiltonian $H_V$ gives the total energy of this particular vortex and
yields its equations of motion. This is true provided that
the background is in the limit of infinite extension, such that its
energy change as the vortex moves, expanding and contracting, is
negligible.

The full Hamiltonian equations of motion 
\begin{equation}
\pard{\vec P}{t} =  -\frac{\delta }{\delta \vec Q} H_V\,,\qquad
\pard{\vec Q} {t} =  \frac{\delta }{\delta \vec P} H_V\,,
\end{equation}
in which $\delta/\delta {\vec Q},\delta/\delta {\vec P} $ formally 
indicate invariant functional derivatives,   
are taking the form (setting $c_s\equiv 1$),
\begin{eqnarray}
\pard{\vec P}t & = & -\nabla_{\vec Q} (\sqrt \gamma M_0)
-q\nabla_{\vec Q}\left(\frac12 a^0_C +a^0_{u}\right)
-\frac12\left( {\vec P}-q{\vec a}\right)^2
\nabla_{\vec Q} \left(\frac{1}{\sqrt \gamma M_0}\right) \nonumber\\
&  & + \,\frac q{\sqrt \gamma M_0}\left(\nabla_{\vec Q} \otimes \vec a
\right)\left( \vec P -q\vec a\right)
+\frac{\partial}{\partial \sigma} \left(\frac{M_0}{\sqrt \gamma}\vec
Q'\right)-\frac12 {\vec
Q}'{}^2\,\nabla_{\vec Q}\left(\frac{M_0}{\sqrt \gamma}\right)\nonumber\\
\label{dPdtHamilton}
\end{eqnarray}
and, of course, $\dot {\vec Q} = (\vec P - q\vec a)/(\sqrt \gamma M_0)$\,.
They are considerably complicated by the fact that we admit 
a dependence of the terms containing $M_0, \gamma$
on the co-ordinates of the line element.
 
The two (vectorial) Hamiltonian equations of motion, being 
of first order in time, can be shown to be
equivalent to the second order Lagrangian equations of motion as 
they follow from the corresponding action (\ref{SV}):
\begin{eqnarray}
\frac{\partial}{\partial t} \left(\frac{M_0}{\sqrt \gamma}\dot {\vec
Q}\right)
-\frac{\partial}{\partial \sigma} \left(\frac{M_0}{\sqrt \gamma}\vec
Q'\right) 
+ \left(1 - \frac 12 \dot {\vec
Q}{}^2\right)\nabla_{\vec Q} (\sqrt \gamma M_0)+\frac12{\vec Q}'{}^2
\nabla_{\vec Q}\,\left(\frac{M_0}{\sqrt \gamma}\right)
\nonumber \hspace*{-3em}\\
= q\left({\vec E} +\dot{\vec Q}\times {\vec B} \right)\,,
\label{Lagrangeeqmotion}
\end{eqnarray}
by making use of the identities 
\begin{eqnarray}
\frac{d \vec P}{dt} = \frac{\partial}{\partial t} 
\left( {\sqrt \gamma}{M_0}\dot {\vec Q}\right)
+ q{\dot{\vec a}} + q \dot {\vec Q}(\nabla_{\vec Q}\otimes
\vec a)\,,\nonumber\\
\dot {\vec Q}\times {\rm rot}\,\vec a= \dot {\vec Q}(\nabla\otimes
\vec a)-(\nabla\otimes\vec a)\dot {\vec Q}\,,\\
\vec E = -\nabla_{\vec Q}\left(\frac 12 a^0_C + a^0_u\right) - \dot
{\vec a}\;,\quad \vec B ={\rm rot}\,\vec a\,.\nonumber
\end{eqnarray}
In these equations, the expression $\nabla\otimes\vec a$
means a second-rank tensor with components $\partial_i a_j$ and a
vector standing to the left or right of this tensor is contracted
with the first or second index, respectively. 

It is instructive to write down the equation of motion in the original
hydrodynamic variables:
\begin{eqnarray}
\frac{\partial}{\partial t} \left({\sqrt \gamma}\ln [\cdots] \dot {\vec
Q}\right)
-\frac{\partial}{\partial \sigma} \left(\frac 1 {\sqrt \gamma}\ln [\cdots]
\vec Q'\right) 
+ \left(1 - \frac 12 \dot {\vec
Q}{}^2\right)\nabla_{\vec Q} (\sqrt \gamma \ln [\cdots])
\nonumber \hspace*{-3em}\\
+\frac12{\vec Q}'{}^2\nabla_{\vec Q}\,\left(\frac 1{\sqrt \gamma}
\ln [\cdots]\right)
= (4\pi/\Gamma_s)  {\vec X}'\times 
\left(\dot{\vec Q}-\vec v_s \right)\,.\label{originaleqmotion} 
\end{eqnarray}  
The logarithm of the self-energy (\ref{Energy}) is abbreviated $\ln [\cdots]$.
The bulk superfluid density $\rho_0$ disappears in this writing.  

\section{Summary}
In this paper, we have developed a general hydrodynamic  
formalism to describe the zero temperature motion of vortices in compressible, 
conventional superfluids, and the fields emanating from and interacting 
with them during this motion.  
The assumptions pertaining to this formalism are as
follows. There exists an entity vortex, whose center of topological 
stability, the singularity of the vortex center, 
can be described by the spacetime embedding 
of the vortex world sheet. The intrinsic property of the vortex is the 
circulation, defined as the line integral of the momentum ${\bm p}$, and 
quantized in units of Planck's quantum of action $h$ 
by the homotopy group index $N_v$. 
In the underlying spacetime,
there exists a conserved particle current density ${\bm j}(x)$, 
whose dual is the field ${\bm H}$ acting on
the vortex. 
The fundamental degrees of freedom in this 
treatment are thus the position of the vortex $X^\mu(\tau,\sigma)$ 
in spacetime, representing the order parameter defect location, 
and the current density ${\bm j}(x)$. In a perfect fluid, these two 
degrees of freedom are connected physically in the 
Magnus force action. This action is given by the linear coupling 
of the vorticity ${\bm \omega}={^*\!{\bm d}}\wedge{\bm p} $ 
and a gauge potential ${\bm b}$, 
whose exterior derivative gives the field strength 
${\bm H}={\bm d}\wedge{\bm b}$. 

In the Galilei invariant, nonrelativistic  fluid we are then dealing with, 
we have further separated the superfluid phase 
into the nonrelativistic part of the
background fluid and a part related to small density and phase
oscillations, namely sound waves. This latter part behaves
`relativistic' in that the equations governing its behaviour have
pseudo-Lorentz invariance, that is, they are Lorentz invariant under 
the replacement $c\rightarrow c_s$. Self-interaction of the vortex
line with these `relativistic' 
phonon excitations gives rise to the hydrodynamic vortex mass.  
In a dense superfluid, the `nonrelativistic' motion 
of the vortex is of relevance, 
because large order parameter modulus (quantum) variations 
take over, long before $|\partial{\vec X}/\partial t|
\lesssim c_s$ is attained. The full canonical structure of vortex
motion obtains  by adding elastic energy in the 
self-action of the vortex string, 
related to the localized self induction approximation.

In the final description of the vortex as a massive, elastic 
string object, we have thus incorporated
parts of the fundamental interaction of the bare vortex with 
the general flow field. The phonon fluctuations in this flow field 
give mass renormalization, whereas the interaction with adjacent 
line segments gives elasticity. The background flow 
potential which remains is well approximated to be coming from 
an incompressible fluid. This procedure yields 
equations of motion in which the role 
of the bare, small scale many-body quantum dynamics of the vortex is 
reduced to give an ultraviolet cutoff parameter in the mass and
elasticity coefficients. 

The local self-action of the laboratory vortex, 
considering only the influence of neighbouring segments on a particular line 
element, has to be modified in a nonlocal manner
if more distant parts of the line are closely approaching each other, 
or when nontrivial topologies of the vortex line, 
knotted structures, are under consideration \cite{SBR}.  
Given that possible vortex configurations are regular and topologically 
trivial in this sense, 
the equations of motion in subsection \ref{Hameqmotion}  
can, for example, 
be used to calculate the Euclidean action of vortex  
tunnelling motions under the influence of external flow fields  
\cite{geo,lammi}. 

   
\bigskip
\begin{center}
{\large\bf Acknowledgements}\end{center}
I thank Nils Schopohl for structurally clarifying discussions.
This research work was supported by
the LGFG Baden-W\"urttemberg. 

\end{document}